\documentclass[3p]{elsarticle}

\usepackage[colorlinks]{hyperref}
\usepackage{lineno}
\usepackage{url}

\usepackage{graphicx}
\usepackage{amsmath}
\usepackage{amssymb}
\usepackage{float}
\usepackage{csquotes}
\usepackage{xcolor}
\usepackage{mathtools}
\usepackage{color}
\usepackage{colortbl}
\usepackage{appendix}
\usepackage{multirow}

\DeclareMathOperator*{\argmax}{arg\,max}

\journal{arXiv}

\begin{document}

\begin{frontmatter}

\title{Limited intelligence and performance-based compensation: An agent-based model of the hidden action problem\tnoteref{mytitlenote}}
\tnotetext[mytitlenote]{This work was supported by funds of the \"Osterreichiscche Nationalbank [Austrian Central Bank Anniversary Fund, project number: 17930].}

\author[mymainaddress]{Patrick Reinwald}
\ead{patrick.reinwald@aau.at}

\author[mymainaddress]{Stephan Leitner\corref{mycorrespondingauthor}}
\cortext[mycorrespondingauthor]{Corresponding author}
\ead{stephan.leitner@aau.at}

\author[mymainaddress]{Friederike Wall}
\ead{friederike.wall@aau.at}

\address[mymainaddress]{University of Klagenfurt, Universit\"atsstra{\ss}e 65-67, 9020 Klagenfurt, Austria}

\begin{abstract}
Models of economic decision makers often include idealized assumptions, such as rationality, perfect foresight, and access to all relevant pieces of information. These assumptions often assure the models' internal validity, but, at the same time, might limit the models' power to explain empirical phenomena. This paper is particularly concerned with the model of the hidden action problem, which proposes an optimal performance-based sharing rule for situations in which a principal assigns a task to an agent, and the action taken to carry out this task is not observable by the principal. We follow the agentization approach and introduce an agent-based version of the hidden action problem, in which some of the idealized assumptions about the principal and the agent are relaxed so that they only have limited information access, are endowed with the ability to gain information, and store it in and retrieve it from their (limited) memory. We follow an evolutionary approach and analyze how the principal's and the agent's decisions affect the sharing rule, task performance, and their utility over time. The results indicate that the optimal sharing rule does not emerge. The principal's utility is relatively robust to variations in intelligence, while the agent's utility is highly sensitive to limitations in intelligence. The principal's behavior appears to be driven by opportunism, as she withholds a premium from the agent to assure the optimal utility for herself. 
\end{abstract}

\end{frontmatter}

%\linenumbers

\section{Introduction}
\label{sec:introduction}

In the history of research on behavioral control -- in particular in the field of economics and from the 1940s onward -- the concept of rational expectations emerged as the dominant paradigm \cite{Muth1961}, which went hand in hand with the extensive development of sophisticated mathematical methods and models with sometimes idealized assumptions about individuals. Up to today, such approaches have kept playing an important role. Most of these models assume intelligent agents who are rational in their behavior, employ optimization methods, usually possess all (or at least most) pieces of information to immediately come up with the optimal solution, and generally make no errors in doing so \cite{Thaler.2000,farmer2009economy}. However, it has already been recognized that technically correct and valid models often lack the power to explain empirical phenomena \cite{Franco2020} and calls to include behavioral insights from other disciplines -- such as cognitive psychology -- have emerged \cite{Royston2013,Hamalainen2013,Leitner.2015b,wall2020agent}. Franco and H\"am\"al\"ainen \cite{Franco2016} point out that increased attention to behavioral aspects becomes prominent whenever scientific fields reach maturity, and argue that, amongst others, this is the case in economics, accounting, and strategic management. As a result of these developments, the field of behavioral operational research (BOR) has emerged \cite{Franco2020}. Two streams can be identified within BOR: First, a line of research that concentrates on the use of operational research methods to model human behavior, and, second, research that analyzes how behavior affects and is affected by model-supported processes \cite{Franco2016}. 

We place our research in the first stream of BOR: In particular, we aim at including more findings related to human intelligence (and limitations thereof) in the well-known model of the hidden action problem introduced by Holmstr\"om \cite{Holmstrom.1979}. This model is based on a principal who assigns a task to an agent. The agent acts on behalf of the principal by making an effort to carry out the task assigned to him, while the principal's role is to provide capital and incentives. The modelled situation is further characterized by information asymmetry in the form of hidden actions. The principal can only observe the task outcome but not the action taken by the agent, which is why the principal can only employ a performance-based compensation mechanism. The model provides this mechanism by coming up with a rule to optimally share the task outcome between the principal and the agent (including optimal risk sharing) \cite{BernardCaillaud.2000,eisenhardt1988agency,Lambert.2001}. The assumptions considered in this model include what Axtell  \cite{Axtell.2007} refers to as the economic sweet spot: Rationality, homogeneity, and equilibrium solutions. We are particularly interested in the conditions of rationality that also include issues related to human intelligence. 

Concepts of human intelligence cover many capabilities that are required for day-to-day activities. Economic models often implicitly follow the idea that decision makers possess these capabilities. It is, for example, often assumed that principals and agents can access information that is relevant for their decision making problems, and can also make sense of it. Keeping in mind that information can take different forms, such as complex texts, multimedia materials, or information that is available by observing the environment, these required capabilities are truly manifold and rich \cite{mccreadie1999trends}. This renders the assumptions implicitly included in economic models nontrivial. Research on human intelligence states that good performances in (related) individual cognitive activities are often positively correlated; this means that if someone performs well in one activity, they are often good at other related activities as well. This is a key finding that has led to the conclusion that human intelligence might be traced back to \textit{one} (condensed) cognitive factor that is often referred to as general intelligence \cite{Sternberg.2018,Chen.2014,KOVACS2019255}. For the context of economic models, this implies that decision makers are usually modelled to be highly intelligent, since they are considered to possess \textit{all} the relevant capabilities to be rational.

There is a stream of model-based research that adopts the opposite approach of deliberately reducing the agents' intelligence to (close to) zero. Models of limited intelligence actors (by limiting their information and/or memory) in the context of financial markets were put forward by Gode and Sunder \cite{gode1993allocative}, Cliff and Bruten \cite{cliff1997zero} and Veryzhenko et al. \cite{veryzhenko2010agent}. In a micro-economic context, Leitner, Rausch, and Behrens \cite{Leitner2017.ejor} and Leitner, Brauneis, and Rausch \cite{Leitner2015.plosone} analyze the effects of limited intelligence in the context of capital budgeting by modelling overconfident decision makers, and principals and agents with limited foresight, respectively. In the context of the hidden action problem, Leitner and Wall \cite{Leitner.2020,wall2020agent} limit the principal's and the agent's respective intelligence by restricting their search capabilities. In the vein of this related research, we transfer the hidden action model into an agent-based model following an agentization procedure introduced, amongst others, by Guerrero and Axtell \cite{Guerrero.2011}, and Leitner and Behrens \cite{Leitner.2015b}. By doing so, we propose an agent-based model of the hidden action problem with principals and agents of limited intelligence. Our model includes principals and agents who suffer from limitations in cognitive functions related to their memory. We focus on the following questions: \textit{How do limitations in the principal's and the agent's memories in the context of the hidden action problem affect (i) the rule to share the task outcome between the principal and the agent, (ii) the agent's effort, and (iii) the principal's and the agent's utilities.}

The remainder of this paper is structured as follows: We discuss the related concepts of human intelligence and put them into the context of economic actors in Sec. \ref{sec:background}. In Sec. \ref{sec:holmstrom}, we introduce the main characteristics of the hidden action model introduced by Holmstr\"om \cite{Holmstrom.1979}. Section \ref{sec:agent-based-model} discusses our approach to limit the principal's and the agent's respective intelligence in more detail, formalizes the agent-based model of the hidden action problem, and discusses the simulation setup. The results are presented in Sec. \ref{sec:results}, while Sec. \ref{sec:discussion} discusses the theoretical and methodological contributions and the managerial implications. Finally, Sec. \ref{sec:conclusion} concludes the paper.

\section{Intelligence and economic agents}
\label{sec:background}

The concept of economic rationality is closely tied to that of human intelligence. The field of economics often distinguishes between rationality at the individual and at the system level. The concept of individual rationality mainly prescribes how humans should make decisions. In particular, it provides guidance on how agents in economic systems, who are given access to relevant pieces of information, should make their decisions so that they achieve their objectives. These objectives include utility maximization, i.e., getting the optimal payoff for a certain or getting the highest expected payoff under the consideration of limits imposed by the conditions and further constraints. Additionally, rational agents are capable of handling the complexity of the decision problem at hand \cite{simon1972theories}. System rationality, on the contrary, is usually concerned with the rationality of the system itself, including the methods and procedures employed in economic systems, such as markets and organizations. System rationality often focuses on the consistency within this functional context. For organizations, system rationality often covers the wider context of methods and procedures to handle conflicting goals  \cite{simon1972theories,godelier2013rationality}.\footnote{For the concepts of rationality (and irrationality), the reader is referred to Einhorn and Hogarth \cite{Einhorn1981}, Forest and Mehier \cite{Forest2001}, Gigerenzer and Selten \cite{Gigerenzer2002}, and Barros \cite{Barros2010} amongst others.} In this paper, we are particularly interested in individual rationality and the limitations thereof caused by limited intelligence.

The cognitive capacity that is required to make rational decisions can, at least to some extent, be measured and is usually termed as intelligence quotient or general intelligence \cite{Thaler.2000,McGrew.2009,Chen.2014}. There is a multiplicity of attempts to characterize human intelligence with a vivid discussion about what cognitive factors are (not) to include \cite{Sternberg.2018}. A well-known \textit{psychometric theory} of intelligence is the Cattell-Horn-Carroll (CHC) theory of cognitive abilities. The CHC theory follows a hierarchical structure with a (condensed) factor for general intelligence on top. At the next level, the CHC theory includes 17 broad cognitive abilities, like fluid intelligence, which stands for the ability to reason and solve problems using new information, long-term storage and retrieval, which stands for the ability to store information in memory and retrieve it later, or quantitative knowledge, which stands for the ability to understand quantitative concepts and to manipulate numerical symbols. At the level of the narrow abilities, the CHC theory includes more than 80 elements that further specify the broad abilities: Fluid reasoning, for example, is further subdivided into different forms of reasoning that can be applied to solve problems using new information, such as inductive, deductive, and quantitative reasoning. If all levels are taken into account, the CHC theory is capable of capturing -- at a rather detailed level -- what human intelligence may be composed of \cite{Flanagan.2014,schneider2018cattell}.\footnote{It is important to note that the theory is at the stage of development, which is why the number and the exact definition of categories are not yet set in stone. For example, Carroll \cite{carroll1993human} found strong evidence for 8 to 10 broad categories, whereas Schneider and McGrew \cite{schneider2018cattell} claim that there might be up to 20 such factors.} 
A widely accepted \textit{cognitive theory} is the triarchic theory of human intelligence, which was introduced by Sternberg \cite{sternberg1984toward}. This theory comprises three components and three subtheories. In particular, Sternberg explains that human intelligence is determined by (i) metacomponents, (ii) performance components, and (iii) knowledge acquisition components and relates to three parts of the triarchic theory, each capturing different tasks. (i) Metacomponents stand for high-order processes used in planning, monitoring, and decision-making, (ii) performance components are the processes used in the execution of a decision-making task, and (iii) knowledge acquisition components mainly capture the processes of learning new things as well as making use of it.  
The contextual subtheory relates human intelligence to the external world of the individual. In particular, it focuses on the processes of adaptation and selection to create an optimal fit between the individual and the environment. This component is mainly concerned with how individuals handle everyday situations, which is why it is also referred to as practical giftedness. 
The creative subtheory takes into account experience with tasks or situations which critically involve the use of intelligence, and generally distinguishes between novel and automated tasks. While the individual has not experienced novel tasks before, automated tasks have already been performed (probably multiple times). This component of human intelligence (also referred to as synthetic giftedness) relates task performance to how well an individual is familiar with a task; it  measures how creative and innovative individuals are when performing tasks. 
The analytical subtheory focuses on intelligence as the internal world of the individual. In particular, it relates intelligence to an individual's academic proficiency and specifies mental mechanisms that lead to more or less intelligent behavior. It covers the ability to systematically break down problems and the capacity to come up with rare solutions. It is also referred to as analytical giftedness \cite{sternberg1984toward}.

It is widely accepted that human cognitive abilities are limited, which -- in turn -- also applies to human intelligence. Taking this into account in the field of economics, it leads to a shift from the assumption of rational decision makers to agents with bounded rationality \cite{arthur1991designing,Thaler.2000,Axtell.2007}.
While there usually is a well-defined way to model agents that are capable of making perfectly rational decisions, there exists no unique way to include limitations in intelligence in economic models. Farmer \cite{Farmer2021} provides an analogy by referring to the multiplicity of ways to limit rationality as the \enquote{wilderness of bounded rationality}. The \enquote{wilderness} is limited by two extreme points -- referred to as gates -- that are located at the two extreme ends: First, the rationality gate, which represents the assumptions related to perfectly rational agents. Second, the zero intelligence gate, which drops all assumptions of perfect rationality and, in consequence, includes decision makers that are not endowed with any intelligence at all. Following Farmer \cite{Farmer2021}, it is necessary to study economic problems from the perspective of both \enquote{gates} because, otherwise, it is likely that the forms of bounded rationality that are taken into account only marginally differ, and that -- in terms of the analogy -- they are located close to each other and around the rationality gate.
Gode and Sunder \cite{gode1993allocative} were among the first to put forward a concept for Zero-Intelligence agents. According to their concept, agents have no intelligence at all, they do not seek or maximize profits, and they are not capable of observing, remembering, or learning specific pieces of information. This concept has been extended by Cliff and Bruten \cite{cliff1997zero}, who do not take zero intelligence for granted, but are interested in the extent of human intelligence that is needed to achieve high-level performance.  With their seminal papers, Gode and Sunder \cite{gode1993allocative} and Cliff and Bruten \cite{cliff1997zero} can be regarded as pioneers in research on zero and/or minimal intelligence agents. Their work has led to rich research on this matter and to further refinements and variations of their concept of zero and/or minimal intelligence agents \cite{veryzhenko2010agent,Chen2008,Ladley2012}. 

\section{Holmstr\"om's hidden action model}
\label{sec:holmstrom}

The hidden action model introduced in Homlstr\"om \cite{Holmstrom.1979} describes the relationship between a principal who assigns a task to an agent. In particular, the principal designs a contract that includes the task that is to be carried out and a rule to share the task outcome, and offers this contract to the agent. If the agent accepts the contract, he makes an effort (often also referred to as action) to carry out the task assigned to him. Together with an exogenous factor, this action generates the task outcome. At the same time, performing the action leads to disutility for the agent. The sharing rule included in the contract defines ex-ante, how the task outcome is to be shared between the principal and the agent. With its rather general formulation, the hidden action model is capable of describing relations between principals and agents in a wide range of areas, including the relation between employer and employee, buyer and supplier, and client and contractor  \cite{Eisenhardt.1989,BernardCaillaud.2000,Leitner.2020,Reinwald.2020}. \\

\begin{figure}
	\centering
	\includegraphics[scale=0.7]{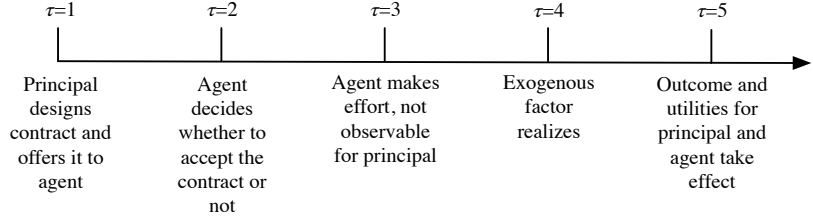}
	\caption{Sequence of events within Holmström's hidden action model}
	\label{fig:timeline}
\end{figure}

The sequence of events within the hidden action model is included in Fig. \ref{fig:timeline}, whereby the main features can be summarized as follows: In $\tau=0$, the principal designs the contract and offers it to the agent who makes his decision about whether or not to accept it in $\tau=1$. In $\tau=2$, the agent selects an effort level $a \in \mathbf{A} \subseteq \mathbb{R}$ from a set of effort levels $\mathbf{A}$ that are feasible to carry out the task. The agent's selected effort level $a$ is \textit{hidden} to the principal, i.e., the principal cannot observe it, because the costs for observing it are prohibitively high or this information is not accessible to her. This results in an information asymmetry regarding the effort level in favor of the agent. The exogenous factor $\theta \sim N(\mu,\theta)$ takes effect in $\tau=3$, it is a random variable that describes the state of nature, which includes, for example, the behavior of suppliers or customers. 
The outcome $x$ materializes in $\tau=4$; it is a function of the agent's effort level $a$ and an exogenous factor $\theta$ and follows the production function 
\begin{equation}
x=x(a,\theta)~.
\end{equation} 
Both the principal and the agent can observe the outcome. However, there is information asymmetry regarding the exogenous factor: The agent can observe the exogenous factor or, given the information about the effort level available to him, deduce it from the outcome. However, the principal cannot observe the exogenous factor and cannot infer it from the outcome. If the principal is aware of the exogenous factor, she can deduce the agent's effort level from the outcome. This, in turn, would render the hidden action problem trivial since all information asymmetry would be resolved, and the principal could pay the agent based on his efforts. As a consequence, the principal can only base the sharing rule on the outcome, so that the agent's share of the outcome is denoted by $s(x)$, whereby $x$ indicates the outcome. 

The principal is risk-neutral. Her utility comes from the generated outcome $x$ minus the agent's compensation $s(x)$, so that 
\begin{equation}
    U_P(x,s(x))=x-s(x)~.
\end{equation}
\noindent The agent is risk-averse and characterized by the utility function
\begin{equation}
    U_A(s(x),a)=V(s(x))-G(a)~,
\end{equation}
where $V(s(x))$ stands for the utility from his share of the outcome and $G(a)$ indicates the disutility of the effort he makes to carry out the task, with $V'>0$ and $x_a \geq 0$.\footnote{The subscript $a$ denotes the partial derivative with respect to $a$.}

Given these characteristics of the hidden action model, the principal's optimization problem to generate the optimal sharing rule is formalized as follows:
\begin{eqnarray}
\label{eq:begin}
\underset{s(\cdot),a}\max E(U_P(x,s(x)))\\
\label{eq:parti}
s.t~ E\{U_A{(s(x),a)\} \geq \bar{U}}\\
\label{eq:incenti}
a \in \underset{a' \in \mathbf{A}}{\argmax} \: E\{U_A(s(x),a')\} 
\end{eqnarray}
Equations (\ref{eq:parti}) and (\ref{eq:incenti}) are constraints that have to be considered by the principal. In particular, Eq. \ref{eq:parti} represents the participation constraint that ensures that the agent gets a minimum utility $\bar{U}$ and therefore accepts the contract. This minimum utility is also referred to as reservation utility and represents agent's best outside option. Equation \ref{eq:incenti} is referred to as the incentive compatibility constraint and aligns the agent's objective (maximize his utility) with the principal's objective. This constraint affects the agent's choice of effort level $a$. The notation \enquote*{arg max} represents the set of all arguments that maximizes the objective function that follows. The notation used in this section is summarized in Tab. \ref{tab:1}. The solution to the hidden action problem is added in \ref{Appendix_B}.

\begin{table}[ht]
\small
	\label{tab:1}
	\begin{tabular}{ll}
		Description& Notation \\
		\noalign{\smallskip}\hline\noalign{\smallskip}
		Effort level (action) & $ a $\\
		Set of feasible actions & $ \mathbf{A} $\\
		Exogenous factor & $ \theta $\\
		Outcome & $ x=x(a,\theta) $\\
		Agent's share of outcome & $ s(x)=x \cdot p $\\
		Principal's utility function & $ U_P(x,s(x)) $ \\
		Agent's utility function & $ U_A(s(x),a) $\\
		Agent's utility from compensation & $ V(s(x)) $\\
		Agent's disutility from effort &$  G(a) $\\
		Agent's reservation utility & $ \bar{U} $\\
		\noalign{\smallskip}\hline
	\end{tabular}
	%}
		\caption{Notation for Holmstr\"om's hidden action model}
\end{table}

\section{An agent-based model of the hidden action problem}
\label{sec:agent-based-model}

\subsection{Limiting the principal's and agent's respective intelligence}

We transfer the hidden action model introduced in Sec. \ref{sec:holmstrom} into an agent based model which, amongst others, allows us to include more realistic behavioral assumptions about the principal's and the agent's respective rationality.\footnote{Further key-features of our approach are that the researcher can analyze rich institutional arrangements and contingencies, and that we can include learning and emergence as well as bridge the micro-macro divide. For a review on the advantages and difficulties related to the approach of agent-based computational economics, the reader is referred to Wall and Leitner \cite{wall2020agent,Leitner2015jomac}. It is important to note that principal-agent models and agentization are two different research paradigms, for further details see Leitner and Behrens \cite{leitner2015fault}.} To do so, we follow the so-called agentization approach, which is a stepwise procedure to systematically relax the assumptions incorporated in economic models to study their robustness in the context of more realistic decision-makers \cite{Guerrero.2011,leitner2015fault}. We follow the vein of Gode and Sunder \cite{gode1993allocative}, Cliff and Bruten \cite{cliff1997zero}, Veryhneko et al. \cite{veryzhenko2010agent}, and Leitner and Wall \cite{Leitner.2020}, and implement limitations in the principal's and agent's respective intelligence by limiting the information available to them.

From Sec. \ref{sec:holmstrom} we know that the key features of the hidden action problem are the information asymmetries regarding the action and the exogenous factor between the principal and the agent. The hidden action model follows the stream of information economics and considers information as imperfect and costly \cite{stiglitz2003information,frieden2010asymmetric,hawkins2010asymmetric}, in contrast to the general equilibrium concepts that view information as given and perfectly known \cite{arrow1973role}. 

\subsubsection{Approach to to transferring the hidden action model}

\subsubsection{Information in the hidden action model}

Since we intend to explicitly limit the information available to the principal and the agent, it is desirable to put the concept of information used in this paper into concrete terms. Concepts of information are manifold, they have their origins in a multiplicity of disciplines, and they are designed with different objectives in mind. Extensive overviews are, for example, provided in McCreadie and Rice \cite{mccreadie1999trends}, Madden \cite{madden2000definition}, Hofkirchner \cite{hofkirchner2009achieve}, and Capurro \cite{capurro2009past}.
At a general level, McCreadie and Rice \cite{mccreadie1999trends} distinguish between information as (i) a resource or commodity, (ii) data in the environment, (iii) representation of knowledge, and (iv) part of the communication process. For economic models, mainly the concept of information as a resource or commodity can be applied. It regards information as entities that can be produced, purchased, replicated, distributed, manipulated, passed along, controlled, traded, or sold, and, thereby, builds the conceptual basis for information asymmetry between parties. We employ the concept of information as a resource or commodity and put the limitations thereof in concrete terms by following Frieden and Hawkins \cite{frieden2010asymmetric}: We denote the information \textit{about a system} by $I$. In the principal-agent context, information $I$ could represent the information available to the principal and the agent. Following McCreadie and Rice's concept of information as a good or a commodity, this information can be produced (e.g., by observation), purchased, replicated, passed along etc. In addition, there is information $J$, which is the most complete information \textit{intrinsic to a system}. Then, any new observation about the system can be modeled as an information flow process $J \rightarrow I$, and the extent to which individuals are informed about a system can be expressed by the distance $J-I$. 
In the principal-agent context described above, information about the outcome $x$ is part of information $I$ for both the principal and the agent, while information about the agent's effort is part of information $I$ for the agent only. 

\begin{table}[ht]
\small
\label{tab:Information}
	\begin{tabular}{lllll}
		\multirow{2}{*}{Type of information} & \multicolumn{2}{l}{Holmstr\"om's model} & \multicolumn{2}{l}{Agent-based model}\\
		    &   Principal   &   Agent   & Principal     & Agent \\
		\noalign{\smallskip}\hline\noalign{\smallskip}
		Agent's utility function                   &  +  &  +  &  +   &  +  \\
    	Agent's reservation utility                &  +  &  +  &  +   &  +  \\       
		Agent's entire action space                &  +  &  +  &  --  &  +  \\
		Action taken by the agent                    &  --  &  +  &  --  &  +  \\
		Production function &  +  &  +  &  +   &  +  \\ 
		Observed outcome                           &  +  &  +  &  +   &  +  \\ 
		Realized exogenous factors                 &  --  &  +  &  --  &  +  \\
		Estimations of exogenous factors           &  n.a.  &  n.a.  &  +   &  --  \\
		\textit{True} expected value of \\exogenous factor &  +  &  +  &  --   &  --  \\ 
		\textit{Learned} expected value of \\
		exogenous factor                           &  n.a.  &  n.a.  &  +  &  +  \\
		\noalign{\smallskip}\hline\noalign{\smallskip}
	\end{tabular}\\
	+ indicates information is available.\\
	-- indicates information is not available.\\
	n.a. indicates information is not considered in the model.
	\caption{Assumptions regarding main pieces of information available in Holmstr\"om's hidden action model and the agent-based model}
\end{table}

\subsubsection{Adaptations to the principal's and agent's intelligence}

In this paper, we are interested in limiting the principal's and the agent's respective information about the environment exclusively. The assumptions about this information in the hidden action model and the agent-based version thereof are summarized in Tab. \ref{tab:Information}. 
In the original hidden action model, both the principal and the agent have full information about the distribution of the exogenous factor. Put into the context of the information concept introduced above, for the principal and the agent, information about the environment is part of the information $J$, and -- since they are perfectly informed -- the distance $J-I$ is equal to zero. Following concepts of human intelligence discussed in Sec. \ref{sec:background}, this indicates that the principal and the agent possess all cognitive abilities to, amongst others, gather, process, store, and retrieve all information about the environment. Since they have full information about the environment, they can compute the \textit{true} expected value for the exogenous variable. Following the logic of the optimization problem introduced in Eqs. \ref{eq:begin}--\ref{eq:incenti}, this enables the principal to immediately come up with the optimal sharing rule. 

In the agent-based model, we limit the principal's and the agent's respective intelligence, so that they no longer possess all relevant intelligence to, for example, gather, process, store, and retrieve information about the environment in the following way:
\begin{itemize}
    \item We limit the principal's information about the environment. She knows that the exogenous variable follows a Normal Distribution, but, due to limited information, she can no longer compute the \textit{true} expected value of the environment. At the same time, we endow her with the ability to estimate the exogenous variable after the outcome has materialized and store the estimations in her memory. 
    \item We also limit the agent's information about the environment so that he cannot compute the \textit{true} expected value of the environment either. The agent is also aware of the exogenous variable's distribution. In contrast to the principal, the agent can observe the realizations of the exogenous variable and store the observations in his memory. 
    \item Since both the principal and the agent learn about the environment, we refer to it as simultaneous and sequential learning. 
    \item The principal's and the agent's respective intelligence is limited in terms of limitations in memory. Once new information is learned, the principal and the agent replace the oldest piece of information stored in their memory by the new estimation and observation, respectively.
    \item Once the information about the exogenous variable is required for decision-making, the principal and the agent retrieve the available information from their memory and compute the expected value based on the information available to them, which is also interpreted as the \textit{perception} of the environment. We refer to it as the \textit{learned} expected value of the exogenous factor.
\end{itemize}

\noindent These adaptations go hand in hand with limitations in human intelligence. Recall, the CHC theory of cognitive abilities \cite{carroll1993human} includes, amongst others, fluid intelligence, which stands for the ability to reason, or the ability to store information in and retrieve it from memory. The limitations in the principal's and agent's respective intelligence described above might be caused by limitations in exactly these cognitive abilities. Taking into account Sternberg's triarchic theory of human intelligence \cite{sternberg1984toward}, the principal and the agent are particularly constrained in the knowledge acquisition components, since they are limited in memory storage and retrieval.

\subsection{The agent-based model of the hidden action problem}

\subsubsection{The principal's and agent's respective utility functions}

\begin{figure}
	\centering
	\includegraphics[scale=0.5]{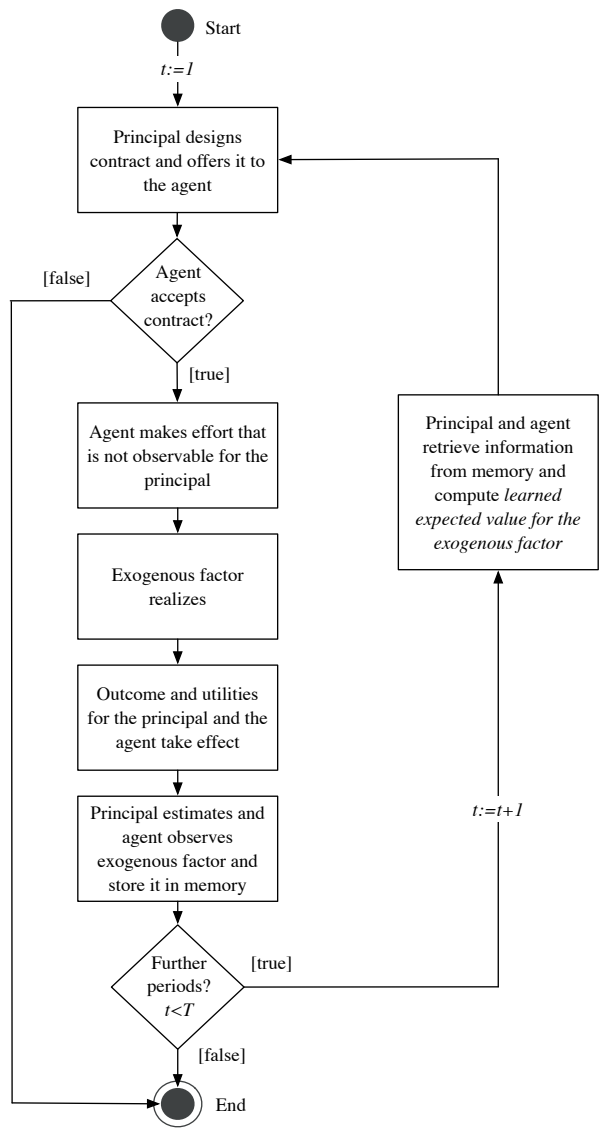}
	\caption{Sequence of events in the agent-based model}
	\label{fig:flow}
\end{figure}

While Holmstr\"om's hidden action model \cite{Holmstrom.1979} is a single-period model, which allows the principal to come up with the optimal sharing rule immediately, the agent-based model is a multi-period model, which includes limitations in the principal's and agent's respective intelligence (and, consequently, information) and a learning mechanism. For the agent-based model, we indicate the time steps by $t=1,...,T$. An overview of the sequence of events is provided in Fig. \ref{fig:flow}.
To account for multiple periods, we reformulate the principal's utility function to
\begin{equation}
\label{utilp_abm}
U_P(x_t,s(x_t))=x_t-s(x_t) ,
\end{equation}
where $x_t$ and $s(x_t)$ denote the outcome and the agent's share of the outcome in $t$. The outcome follows the production function 
\begin{equation}
\label{eq:outcome_abm}
    x_t=a_t + \theta_t ,
\end{equation}
where $\theta_t\sim N(\mu,\sigma)$ indicates the exogenous factor that takes effect in $t$, and $a_t$ indicates the agent's effort in $t$. 
The sharing rule includes $\rho_t$, which indicates the premium parameter in $t$, and is formalized by 
\begin{equation}
\label{remu_abm}
s(x_t)=x_t\cdot\rho_t~.
\end{equation}
The principal can adjust the premium parameter $\rho_t$ in every timestep, i.e., whenever the principal designs a new contract, she has a specific effort level $\tilde{a}_t$ in mind that she wants the agent to make. We refer to $\tilde{a}_t$ as incited effort. In every $t$, the principal can adapt the premium parameter to provide the agent with incentives to make this effort. Details on the computation of the incited effort and the premium parameter are provided in Sec. \ref{sec:premiumparam}. 
 
The agent's CARA utility function is formalized by
\begin{equation}
\label{utila_abm}
U_A(s(x_t),a_t)= \underbrace{\dfrac{1-e^{-\eta\cdot s(x_t)}}{\eta}}_{V(s(x_t))}-\underbrace{\dfrac{a_t^2}{2}}_{G(a_t)},
\end{equation}
where $\eta$ denotes the Arrow-Pratt measure of risk-aversion \cite{arrow1973role}. The notation used in the agent-based model is summarized in Tab. \ref{tab:abm}.

\begin{table}
\small
\label{tab:abm}       
\begin{tabular}{ll}
		Description& Notation \\
		\noalign{\smallskip}\hline\noalign{\smallskip}
		Timesteps                                       & $ t $ \\
		Principal's utility                             & $ U_P $ \\
		Agent's utility                                 & $ U_A $ \\
		Agent's Arrow-Pratt measure of risk-aversion    & $ \eta $ \\
		Outcome in $t$                                  & $x_t=a_t+\theta_t$ \\
		Agent's effort in $t$ & $ a_t $\\
		Incited effort in $t$ & $ \tilde{a}_t $\\
		Exogenous variable in $t$ & $\theta_t$\\
		Principal's expected outcome for effort level $a'$ in $t$   & $\tilde{x}_{Pt}(a') $\\
		Agent's expected outcome for effort level $a'$ in $t$       & $\tilde{x}_{At}(a') $\\
		Agent's share of the outcome in $t$                     & $ s(x_t)=x_t \cdot p_t $ \\
		Premium parameter included in the contract in $t$   & $\rho_t$\\
		Premium parameter for effort level $a'$ in $t$      & $\rho_t(a')$\\
		Principal's set of all feasible actions in $t$      & $ \mathbf{A}_{Pt} $ \\
		Principal's candidates for incited effort in $t$    & $ \mathbf{\tilde{A}}_{Pt} $ \\
		Agent's set of all feasible actions in $t$          & $ \mathbf{A}_{At} $ \\
	    Principal's memory & $m_P$\\
		Agent's memory & $m_A$\\
		Estimations of the exogenous factors available to the principal in $t$      &   $\mathbf{\Theta}_{Pt}$ \\
		Observations of the exogenous factors available to the agent in $t$      &   $\mathbf{\Theta}_{At}$ \\
		Principal's \textit{learned} expectation of the exogenous factor in $t$ & $\tilde{\theta}_{Pt}$ \\
		Agent's \textit{learned} expectation of the exogenous factor in $t$ & $\tilde{\theta}_{At}$ \\
		\noalign{\smallskip}\hline
	\end{tabular}
	\caption{Notation for the agent-based model}
\end{table}

\subsubsection{Simultaneous and sequential learning model}

The principal and the agent both dispose of an individual memory $m_P$ and $m_A$, respectively. The higher $m_P$ and $m_A$, the more estimations and observations of the exogenous factor the principal and the agent can store in their memory.
Due to different states of information, the agent's actual effort $a_t$ might deviate from the incited effort $\tilde{a}_t$. However, the principal has no information about the actual effort $a_t$ and, therefore, bases her estimation of the exogenous factor in $t$ on the incited effort. She computes her estimation according to
	\begin{equation}
	\label{calcThetaPrincipal}
	\tilde{\theta}_t=x_t- \tilde{a}_t ~.
	\end{equation}
The agent, in contrast, has information about the actual effort $a_t$ he made and, therefore, can compute the actual exogenous factor in $t$ according to
		\begin{equation}
	\label{calcThetaAgent}
	\theta_t= x_t-a_t~.
	\end{equation}
Both the principal and the agent store their estimations and observations in their memories.  Once their capacities, $m_P$ for the principal and $m_A$ for the agent, are reached, the oldest piece of information is replaced by the new estimation or observation. 

Once the simulation moves on to the next timestep, the principal and the agent update the \textit{learned} expected value for the exogenous factor by averaging all privately stored estimations/observations. To do so, they retrieve the information from their memories. Let us denote the estimations of the exogenous factor available to the principal in $t$ by
\begin{equation}
\label{eq:memory_principal}
    \mathbf{\Theta}_{Pt}=
    \begin{cases}
    [\tilde{\theta}_1, \dots, \tilde{\theta}_{t-1}]           & \text{if~}m_P=\infty~,\\
    [\tilde{\theta}_{t-{m_P}}, \dots, \tilde{\theta}_{t-1}]   & \text{if~}m_P<\infty \text{~and~} t\geq m_P~,\\
    [\tilde{\theta}_1, \dots, \tilde{\theta}_{t-1}]           &\text{if~}m_P<\infty \text{~and~} t< m_P~.\\
    \end{cases}
\end{equation}
For the agent, we denote the observations of the exogenous factor available in $t$ by
\begin{equation}
\label{eq:memory_agent}
    \mathbf{\Theta}_{At}=
    \begin{cases}
    [{\theta}_1, \dots, {\theta}_{t-1}]           & \text{if~}m_A=\infty\\
    [{\theta}_{t-{m_A}}, \dots, {\theta}_{t-1}]   & \text{if~}m_A<\infty \text{~and~} t\geq m_A\\
    [{\theta}_1, \dots, {\theta}_{t-1}]           &\text{if~}m_A<\infty \text{~and~} t< m_A\\
    \end{cases}
\end{equation}
The principal and the agent compute their \textit{learned} expected value of the exogenous factor in $t$ as the mean $\varnothing(\cdot)$ of the information available to them: For the principal, the  \textit{learned} expected value of the exogenous factor is $\hat{\theta}_{Pt}=\varnothing(\mathbf{\Theta}_{Pt})$, while  for the agent it is computed according to $\hat{\theta}_{At}=\varnothing(\mathbf{\Theta}_{At})$. Note that the \textit{learned} expected exogenous factor can also be interpreted as the principal's and the agent's perceptions of the environment, since it represents how they truly regard the environment in that timestep. Then, their intelligence can be interpreted as a moderator for their perception \cite{Duncan1972}. 

\subsubsection{The principal's and agent's decisions}
\label{sec:premiumparam}

The principal's and the agent's decisions revolve around the selection of actions and the computation of the corresponding premium parameters. We denote the set of feasible actions from the perspective of the principal and the agent by $\mathbf{A}_{Pt}$ and $\mathbf{A}_{At}$, respectively. The lower boundary of these spaces is defined by the participation constraint and the upper boundary is given by the incentive compatibility constraint \cite{Holmstrom.1979,Leitner.2020}. Recall that the computation of the two constraints includes the expectation about the exogenous factor (see Eqs. \ref{eq:parti} and \ref{eq:incenti}). Thus, if the principal and the agent have the same (different) expectations about the environment, $\mathbf{A}_{Pt}$ and $\mathbf{A}_{At}$ perfectly coincide (are different).
	
\paragraph{The principal's decision}	
	
In every timestep, the principal can adapt the premium parameter in the contract. To do so, she randomly discovers two alternative effort levels in the action space $\mathbf{A}_{Pt}$, $\hat{a}_1$ and $\hat{a}_2$, which together with the incited effort of the previous period $\tilde{a}_{t-1}$ serve as candidates for the effort she wants the agent to make in period $t$. 
Let us denote the set of candidates for the incited effort in $t$ by $\mathbf{\tilde{A}}_{Pt}=[\hat{a}_1, \hat{a}_2, \tilde{a}_{t-1}]$. 
In line with Eq. \ref{eq:outcome_abm}, the principal computes the expected outcome for all alternatives $a' \in \mathbf{\tilde{A}}_{Pt}$ according to
 \begin{equation}
     \tilde{x}_{Pt}(a') = a' + \hat{\theta}_{Pt}~.
 \end{equation}
The principal also computes the premium parameters for all candidate effort levels $a' \in \mathbf{\tilde{A}}_{Pt}$ according to 
\begin{equation}
\rho_t(a') = \max_{\rho\in[0,1]} U_P(\tilde{x}_{Pt}(a'),s(\tilde{x}_{Pt}(a')))~,
\end{equation}
and finally selects the candidate which maximizes her utility as the incited effort for the period $t$ according to
\begin{equation}
    \tilde{a}_{t} = \max_{a' \in \mathbf{\tilde{A}}_{Pt}} U_P(\tilde{x}_{Pt}(a'),s(\tilde{x}_{Pt}(a')))
\end{equation}
\noindent Together with the task that is to be carried out (which is the same throughout all time steps), the corresponding premium parameter $\rho_t := \rho_t( \tilde{a}_{t})$ is the main element of the contract that is offered to the agent.

\paragraph{The agent's decision}
In every timestep, the agent makes two decisions. First, he decides whether to accept or reject the contract offered by the principal. In particular, if the utility of the offered contract exceeds the reservation utility so that the agent accepts the contract. To make this decision, the agent computes the effort that maximizes his utility given the offered contract according to
\begin{equation}
    a^{\ast}_t = \max_{a'\in \mathbf{A}_{At}} U_A (s(\tilde{x}_{At}(a')), a')~,
\end{equation}
where $\tilde{x}_{At}(a') = a' + \tilde{\theta}_{At}$. If the utility of this effort level exceeds or is equal to the agent's reservation utility,
\begin{equation}
    U_A (s(x_t), a^{\ast}_t) \geq \overline{U}~,
\end{equation}
the agent accepts the contract and makes this effort in $t$, so that $a_t := a^{\ast}_t$.\footnote{Please note that, without loss of generality, we normalize the reservation utility to zero during the simulation experiments, and make sure that the agent accepts the contract in all cases. }

\subsection{Parameter settings and simulation experiments}
\label{sec:para}

We analyze scenarios with four different values for the principal's memory ($m_P= \{1,3,5,\infty\}$) and the agent's memory ($m_A=\{1,3,5,\infty\}$), whereby this parameter can be interpreted so that the poorer the memory, the lower the intelligence. In addition, we take into account the turbulence of the environment: Recall, the exogenous variable follows a Normal Distribution, which allows us to control the turbulence via the standard deviation. In particular, we set the mean of the Normal Distribution to zero and define its standard deviation relative to the optimal outcome $x^{\ast}$ of Holmstr\"om's hidden action model (see \cite{Holmstrom.1979} and \ref{Appendix_B}), so that $\sigma=\{0.05x^{\ast}, 0.25x^{\ast}, 0.45x^{\ast}\}$). Variations in these parameters result in $4 \cdot 4 \cdot 3 = 48$ scenarios. 
All other parameters are kept constant during the simulation runs. Our analysis focuses on the first $T=20$ timesteps in every simulation round, and every scenario is repeated $R=700$ times.\footnote{We follow the approach proposed in Lorscheid et al. \cite{lorscheid2012opening} and select the number of simulation rounds on the basis of the coefficient of variation. For our settings, this measure stabilizes at around 700 repetitions.} An overview of the parameters considered in this simulation study is provided in Tab. \ref{tab:key_para}.

\begin{table}[ht]
\small
	\label{tab:key_para}   
	\begin{tabular}{lll}
		Parameter & Notation & Values \\
		\noalign{\smallskip}\hline\noalign{\smallskip}
		\textbf{Subject to variation}\\
		Principal's memory & $ m_P $ & $\{1,3,5,\infty\}$\\
		Agent's memory & $ m_A $  & $\{1,3,5,\infty\}$\\
		Exogenous factor: standard deviation & $ \sigma $ & $\{0.05x^{\ast},0.25x^{\ast}, 0.45x^{\ast}\}$\\
		\textbf{Constants}\\
	    Exogenous factor: mean & $ \mu $ & $ 0$\\
		Agent's Arrow-Pratt measure & $ \eta $ & $0.5$\\
		Observation period  &   $T$ &   $20$ \\
		Simulation rounds  &   $R$ &   $700$ \\
		\noalign{\smallskip}\hline
	\end{tabular}
		\caption{Simulation parameters}
\end{table}

\section{Results}
\label{sec:results}

The results are organized in three sections. First, Sec. \ref{sec:results-premiumparameter} analyzes the effect of limitations in intelligence on the premium parameter. Second, Sec. \ref{sec:results-performance} provides an analysis of the effort made by the agent. Third, Sec. \ref{sec:results-utility} focuses on the dynamics within the agent-based model and puts particular emphasis on the principal's and the agent's utilities that result from the choices related to the premium parameter and the effort.  
The results presented in Secs. \ref{sec:results-premiumparameter}--\ref{sec:results-utility} focus on two levels of memory for both the principal and the agent, and three levels of environmental turbulence:
\begin{itemize}
    \item For the principal's and the agent's memories, we present the results for low and high memory, so that $m_P=\{1,5\}$ and $m_A=\{1,5\}$, respectively. Recall, a memory of $1$ indicates a minimally intelligent decision maker, and the higher the memory, the higher the intelligence. 
    \item For the environment, we consider the cases of stable, moderately turbulent, and turbulent environments and set the parameters to $\sigma=0.05x^{\ast}$, $0.25x^{\ast}$, and $0.45x^{\ast}$, respectively.
\end{itemize}
We focus on these scenarios, since the patterns for the remaining scenarios of medium and unlimited memory, i.e., for $m_P=\{3,\infty\}$ and $m_A=\{3,\infty\}$, are identical to those of the scenarios in our focus. The full set of results is added in \ref{Appendix_A}.

\subsection{Effects on the premium parameter}
\label{sec:results-premiumparameter}

This section focuses on the analysis of the premium parameter that emerges over time. Recall, the principal can adapt the premium parameter in every timestep to provide the agent with incentives to take the action the principal wants him to take. As introduced in Eq. \ref{remu_abm}, the premium parameter defines the agent's share of the task outcome. We denote the premium parameter that is effective in the period $t$ and simulation run $r$ by $\rho_{tr}$ and the premium parameter effective in the optimal solution by $\rho^{\ast}$ (see  \ref{Appendix_B}). We compute the normalized average premium parameter in every timestep according to 

\begin{equation}
    \tilde{\rho}_t=\frac{1}{R} \sum_{r=1}^{R} \frac{\rho_{tr}}{\rho^{\ast}} ~.
\end{equation}

\begin{figure}[ht]
	\centering
	\includegraphics[width=0.6\textwidth]{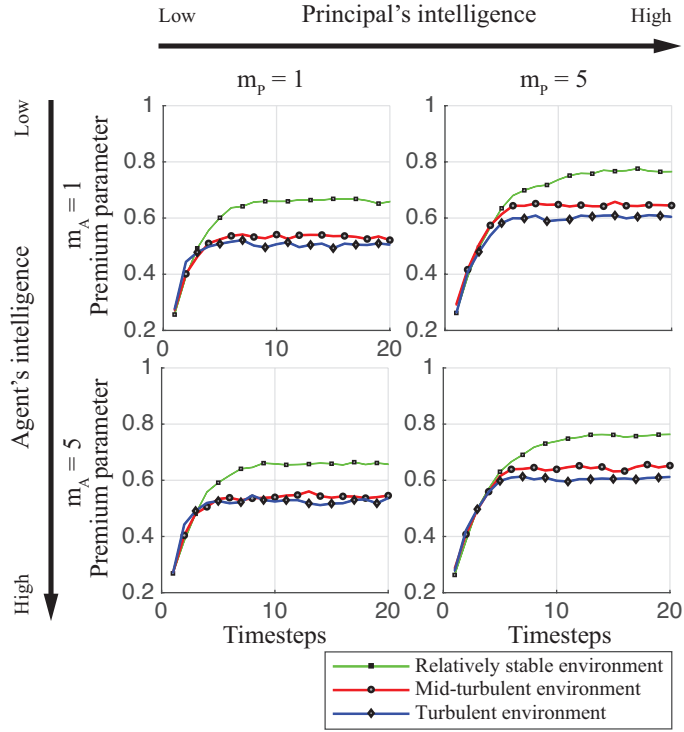}
	\caption{Normalized average premium parameter for selected scenarios. Shaded areas represent confidence intervals at the 99\%-level}
	\label{fig:premiump_selection}
\end{figure}

The normalized average premium parameters are plotted in Fig. \ref{fig:premiump_selection}, which is composed of four subplots representing all combinations of the considered cases for the principal's and the agent's memories. Every subplot contains three lines that represent the three levels of environmental turbulence. The full set of results for the premium parameter is included in \ref{Appendix_A}. 
The results indicate that, in general, the premium parameter decreases with higher environmental turbulence.  Please note that the turbulence \textit{perceived} by the principal and the agent is also affected by their memories. Recall the principal's and the agent's memories introduced in Eqs. \ref{eq:memory_principal} and \ref{eq:memory_agent}, respectively: The higher the memory, the more estimations/observations of the exogenous variable are taken into account when the learned expectation about the exogenous variable is computed. Thus, more memory translates into more precise expectations and, as a consequence, less perceived turbulence over time.  
In all subplots, the premium parameter is the highest (lowest) in stable (turbulent) environments. 
Taking a closer look at the effects of the principal's and the agent's respective intelligence in this context reveals that irrespective of the agent's memory, more intelligent (or better informed) principals will set higher premium parameters. For increases in the agent's memory, in contrast, no significant effects on the premium parameter can be observed. This is in line with expectation, since it is the principal who \textit{designs} the contract (and, thereby, fixes the premium parameter).

\subsection{Effects on the agent's effort}
\label{sec:results-performance}

\begin{figure}[ht]
	\centering
	\includegraphics[width=0.6\textwidth]{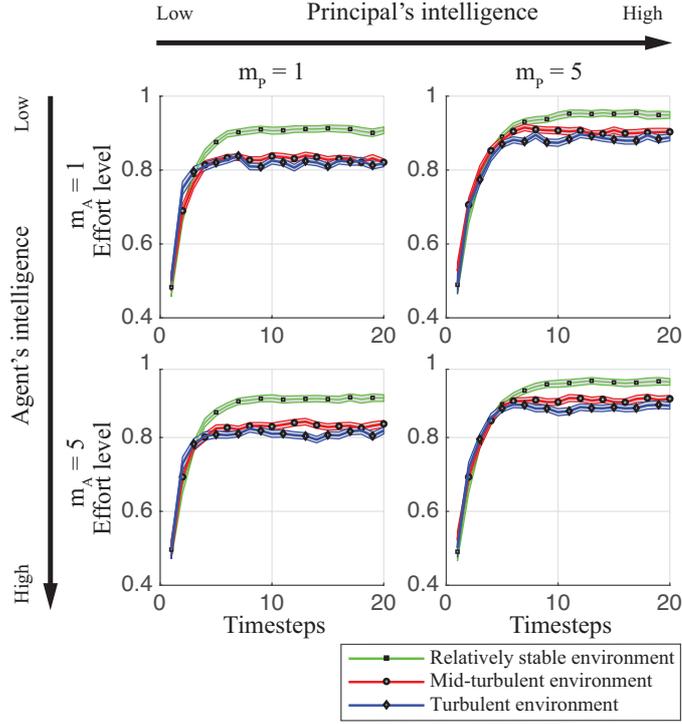}
	\caption{Normalized average effort for selected scenarios. Shaded areas represent confidence intervals at the 99\%-level}
	\label{fig:a_A_selection}
\end{figure}

This section focuses on the analysis of how the agent's behavior responds to the premium parameter set by the principal.
To do so, we report the normalized average actual effort level in every timestep $t$, which is calculated as follows:

\begin{equation}
\label{eq:performanceIndi}
\tilde{a}_t = \dfrac{1}{R}\sum_{r=1}^{r=R}\dfrac{a_{t r}}{a^{\ast}}~,
\end{equation}
where $ a_{tr}$ indicates the effort made by the agent in timestep $t$ and simulation run $r$, and $a^{\ast}$ stands for the optimal level of effort that is suggested by Holmstr\"om's model \cite{Holmstrom.1979} (see \ref{Appendix_B}). Please note that the normalized average effort is also a proxy of the task outcome at the macro level, as the outcome is a function of the agent's effort and the exogenous variable (see Eq. \ref{eq:outcome_abm}).
The agent's normalized average effort is plotted in Fig. \ref{fig:a_A_selection}, where subplots indicate different settings for the principal's and the agent's memories, and the lines within the subplots stand for different levels of environmental turbulence. The full set of results for the agent's effort is included in \ref{Appendix_A}. 

The results indicate that the agent responds to the incentives provided by the principal, since higher premium parameters result in more effort. In Sec. \ref{sec:results-premiumparameter} it was discussed that the premium parameter increases as the environmental turbulence decreases, which also holds for the agent's effort. It is remarkable that -- for scenarios with relatively intelligent principals and relatively stable environments -- the effort almost reaches the level of the optimal effort suggested 
by Holmstr\"om's hidden action model. However, even though the premium parameters are significantly different in cases with relatively intelligent principals and moderate to high environmental turbulence (see Fig. \ref{fig:premiump_selection}), we can not generally observe any significant differences in the agent's effort in these scenarios. 

\subsection{Effects on the principal's and agent's utilities}
\label{sec:results-utility}

This section focuses on the analysis of the effects of limited intelligence on the principal's and agent's utilities. To do so, we report the normalized averaged utilities. 
For the principal, we denote the utility experienced in time step $t$ and simulation run $r$ by $U_{Prt}$ (see Eq. \ref{utilp_abm}) and normalize it by the utility that can be achieved with the solution of Holmstr\"om's hidden action model, which we denote by $U^{\ast}_P$ (for its computation see \ref{Appendix_B}). Then, the principal's normalized average utility at time $t$ is computed according to 
\begin{equation}
\label{eq:up_results}
    \tilde{U}_{Pt} = \frac{1}{R} \sum_{r=1}^{R} \frac{U_{Prt}}{U^{\ast}_{P}}~.
\end{equation}
Similarly, we denote the agent's utility (see Eq. \ref{utila_abm}) in timestep $t$ and simulation run $r$ and the utility following Holmstr\"om's model by $U_{Art}$ and $U^{\ast}_A$, respectively. We compute the agent's normalized average utility at time $t$ by

\begin{equation}
\label{eq:ua_results}
    \tilde{U}_{At} = \frac{1}{R} \sum_{r=1}^{R} \frac{U_{Art}}{U^{\ast}_{A}}~.
\end{equation}

\begin{figure}[ht]
	\centering
	\includegraphics[width=0.6\textwidth]{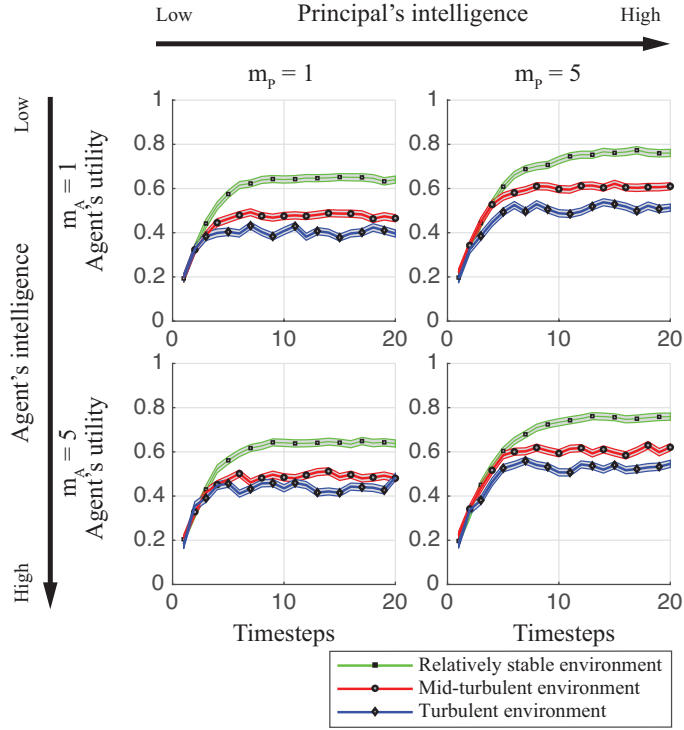}
	\caption{Agent's normalized average utility for selected scenarios. Shaded areas represent confidence intervals at the 99\%-level}
	\label{fig:UA_selection}
\end{figure}

\begin{figure}[ht]
	\centering
	\includegraphics[width=0.6\textwidth]{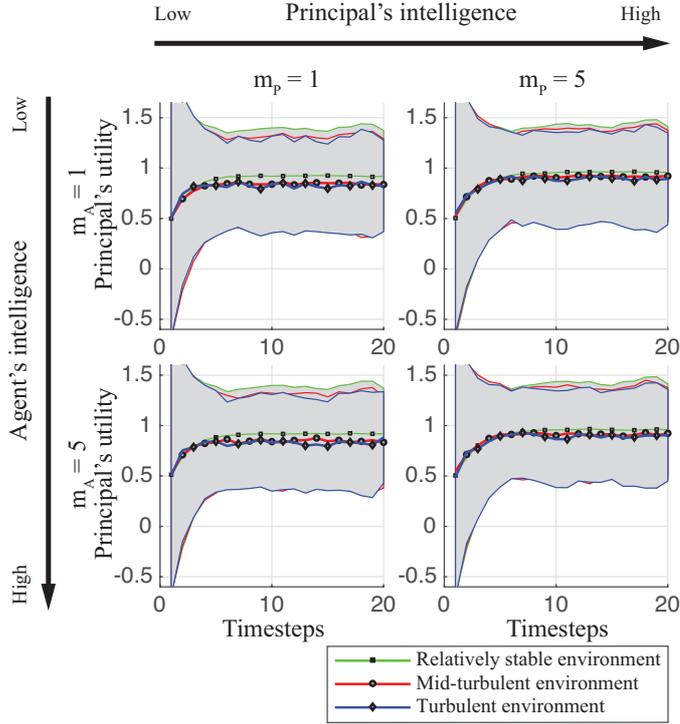}
	\caption{Principal's normalized average utility for selected scenarios. Shaded areas represent confidence intervals at the 99\%-level}
	\label{fig:UP_selection}
\end{figure}

The agent's and the principal's normalized averaged utilities are presented in Figs.  \ref{fig:UA_selection} and  \ref{fig:UP_selection}, respectively. Each figure is composed of 4 subplots representing different parameters for the principal's and the agent's memories. Within every subplot, the lines indicate three levels of environmental turbulence. For the agent's utility, we report the Euclidean Distance between two time series if either the principal's or the agent's respective intelligence is increased in Tab. \ref{tab:influence}, and we also indicate whether the compared time series are significantly different or not.\footnote{We use the standard measure of the Euclidean Distance: Let two time series, each with 20 values, be denoted by $\mathbf{X}=[x_1, \dots, x_{20}]$ and $\mathbf{Y}=[y_1, \dots, y_{20}]$. We then compute the Euclidean Distance between the two time series by $\sqrt{\sum_{i=1}^{20} (x_i-y_i)^2}$.} All further plots for the principal's and the agent's utilities are included in \ref{Appendix_A}. 

\begin{table}[ht]
\small
	\label{tab:influence}   
	\begin{tabular}{p{2.5cm}p{1.3cm}p{1.4cm}p{1.4cm}p{1.4cm}p{1.4cm}}
         &   & \multicolumn{2}{p{2.9cm}}{Increases in P's \newline intelligence} & \multicolumn{2}{p{2.9cm}}{Increases in A's \newline intelligence}    \\
         \noalign{\smallskip}	\cline{2-6}\noalign{\smallskip}
        \multirow{2}{3cm}{Environment}& $m_P$   & $1\rightarrow5$   &  $1\rightarrow5$ & $1$ & $5$ \\
        & $m_A$ &  $1$  &  $5$ & $1\rightarrow5$ & $1\rightarrow5$ \\
		\noalign{\smallskip}\hline\noalign{\smallskip}
		\multirow{2}{3cm}{Relatively stable} 
		        & Distance          & $0.3931$    & $0.3996$    & $0.0422$    & $0.0432$  \\
		        & Sign.      & **       &   **     & n.s.      & n.s. \\
		\noalign{\smallskip}\hline\noalign{\smallskip}
				\multirow{2}{3cm}{Mid-turbulent} 
				& Distance          & $0.5134$    & $0.4757$    & $0.0800$    & $0.0631$  \\
		        & Sign.      & **       &   **     & n.s.      & n.s. \\
		\noalign{\smallskip}\hline\noalign{\smallskip}
				\multirow{2}{3cm}{Turbulent} 
				& Distance          & $0.4422$    & $0.3778$    & $0.1876$    & $0.1134$  \\
		        &   Sign.    & **       &   **     & n.s.      & n.s. \\
		\noalign{\smallskip}\hline\noalign{\smallskip}
	\end{tabular}\\
	P=principal, A=agent, Sign.=significance.\\
	$1\rightarrow 5$ indicates that the memory is increased from $1$ to $5$.\\
	** indicates significance at the $99\%$-level.\\
	n.s. indicates no significant difference.
	\caption{Influence of intelligence on the agent's utility}
\end{table}

The results in Fig. \ref{fig:UA_selection} and Tab. \ref{tab:influence} suggest that limitations in intelligence significantly decrease the agent's utility in all cases.\footnote{Please note that by the construction of the performance measures in Eqs. \ref{eq:up_results} and \ref{eq:ua_results}, the utilities achievable with the solution proposed by Holmstr\"om's model \cite{Holmstrom.1979} take a value of $1$.} In the worst case, the agent loses around $60\%$ of his utility, while in the best case, the lost utility amounts to a value of only around $20\%$. The agent's utility follows the pattern of the premium parameter presented in Fig. \ref{fig:premiump_selection}. First, increases in (perceived) environmental turbulence decrease the agent's utility. Second, if the principal is better informed about the environment, the agent's utility increases as well. Third, no significant effects on utility can be observed if the extent of information available to the agent increases. These are interesting findings, since the agent's utility is composed of two components (see Eq. \ref{utila_abm}): First, the utility of compensation which is mainly in the sphere of control of the principal, who sets the premium parameter. Second, the disutility of making an effort, which, on the contrary, is in the agent's sphere of control, since he autonomously decides about the action. The results included in Fig. \ref{fig:UP_selection} indicate that the principal attains a near-optimal utility in all scenarios. What is particularly surprising is that -- irrespective of the environmental turbulence, the principal's and the agent's memories -- the principal's utility is at the same level in all scenarios. This means that the principal's utility appears to be robust to limitations in her or the agent's intelligence. 

\section{Discussion}
\label{sec:discussion}

\subsection{Theoretical contributions}

\begin{figure}[ht]
	\centering
	\includegraphics[width=0.7\textwidth]{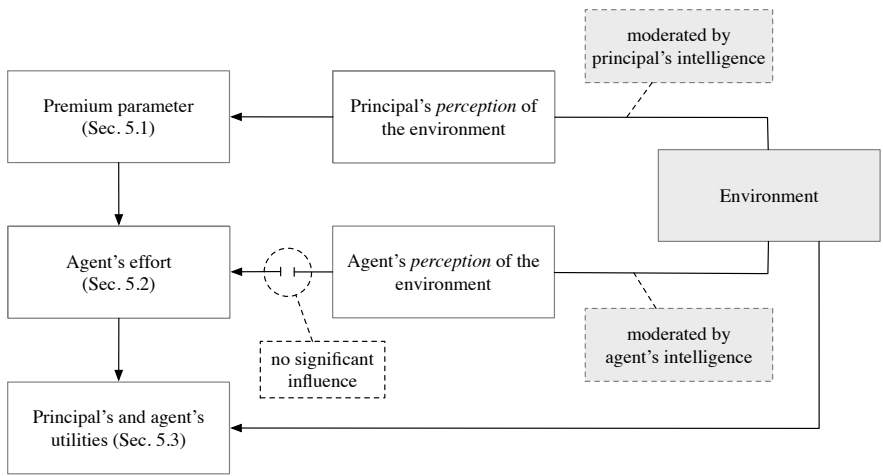}
	\caption{Framework for limited intelligence in hidden action situations}
	\label{fig:discussion}
\end{figure}

The results presented in Sec. \ref{sec:results} allow to put the effects of limitations in the principal's and agent's respective intelligence and environmental turbulence in the following framework (see Fig. \ref{fig:discussion}): Recall, the principal and the agent estimate and observe the exogenous variable in every timestep, respectively (see Eqs. \ref{calcThetaPrincipal} and \ref{calcThetaAgent}). Then, they store their estimations/observations in their memories. Limitations in their respective intelligence take effect in the form of limitations in their memories (see Sec. \ref{sec:background}). The more intelligent the principal and the agent are, the more pieces of information stored in their memories are taken into account when computing the \textit{learned} expectation of the exogenous variable (see Eqs. \ref{eq:memory_principal} and \ref{eq:memory_agent}). Thus, the principal's and the agent's \textit{perceptions} of the environment are moderated by their respective intelligence. The principal's perception affects her choice of the premium parameter (see Sec. \ref{sec:results-premiumparameter}). One might expect that the agent's perception of the environment affects his choice of effort, but we could not observe significant influences for this relation (see Sec. \ref{sec:results-performance}). Together with the actual environment, the principal's and the agent's decisions about the premium parameter and the effort, respectively, affect their utilities (see Sec. \ref{sec:results-utility}). The analysis in this paper focuses particularly on the effects of turbulence in the environment and limitations in the principal's and agent's respective intelligence in this framework (which is indicated by the gray boxes in Fig. \ref{fig:discussion}).

\paragraph{Discussion of results related to the premium parameter}

The main findings presented in Sec. \ref{sec:results-premiumparameter} are that (i) the value of the premium parameter decreases with environmental turbulence and (ii) increases with the principal's intelligence. This means that the agent's share of the outcome is relatively small (large) if the environment is relatively turbulent (stable). However, if the turbulence is high, the agent cannot easily control the outcome, which puts his compensation at risk \cite{Miceli2000}. Accordingly, the finding that the (perceived) turbulence decreases the premium parameter is (in part) contrary to the risk premium hypothesis. This hypothesis states that the principal will have to increase the risk-averse agent's total compensation in turbulent environments to protect him from risk \cite{eisenhardt1988agency,umanath1993impact,Burns2001}. Counter-intuitively, the results presented in Sec. \ref{sec:results-premiumparameter} indicate the contrary. Even though the risk-bearing is originally within the principal's role \cite{Fama1983}, she shifts a part of the risk to the agent.

This finding not only contradicts the risk premium hypothesis, but is also in contrast to previous research on task programmability \cite{Stroh1996} and the predictions of classical organization theory \cite{Thompson1967}: Stroh et al. \cite{Stroh1996}, and Sung and Choi \cite{Sung2012} link environmental turbulence with task programmability by arguing that more (less) turbulence can be interpreted in terms of less (more) programmable tasks, since task situations and problems change more frequently in turbulent environments. Furthermore, it is often suggested that task programmability is negatively correlated with the magnitude of variable compensation components \cite{Gomez-Mejia1992}. This translates into the expectation that more environmental turbulence (i.e., less task programmability) should be linked to higher proportions of variable compensation, and vice versa \cite{Stroh1996,Gerhart1990}. This expectation is also in line with classical organization theory \cite{Thompson1967}, which predicts that organizations that operate in turbulent environments rely more heavily on variable compensation to ensure the appropriate behaviors of their agents. Our findings do not support this expectation but can be linked to the argumentation provided in Smith \cite{Smith1984}. There are not just drawbacks but also advantages of turbulent environments \cite{Eisenhardt.1989}, and an analogy of agents who are hired in risky environments and \enquote*{unfair lotteries} can be established: The \textit{average} pay is not necessarily high, but there is a chance to receive a high compensation, even though perhaps this chance is rather small (see also \cite{Miceli2000}).

\paragraph{Discussion of results related to the agent's effort}

The results presented in Sec. \ref{sec:results-performance} indicate that the agent's effort follows the patterns observed for the premium parameters. The effort decreases with environmental turbulence, so that the agent makes more (less) effort in relatively stable (turbulent) environments. If the premium parameters are set by a more intelligent principal, the agent makes significantly more effort. This relation is a fundamental assumption in economic contexts and supported by experimental \cite{Dickinson1999,Takahashi2016} and field experiments \cite{Banker1996,Lazear2000}. %The results, thus, indicate that these findings also hold for principals and agents with limited intelligence, and we could not observe any crowding-out effects, like it is often suggested in the field of Social Psychology \cite{Deci1971,Frey2001}.

Since the agent's average normalized effort is also a proxy for how well the organization performs, these observations can be related to research on environmental turbulence and firm performance. Environmental uncertainty \cite{milliken1987three,chen2005impacts} and environmental dynamism \cite{aldrich2008organizations} increase the difficulty of organizational decision making and significantly affect organization performance \cite{reinwald2021effects}.  This is in line with the results presented in Yu et al. \cite{Yu2016}, who find that (perceived) environmental uncertainty affects the identification of competitors, so that firms identify more competitors if the environment is certain. This directly translates into a more competitive advantage and, as a consequence, a better (worse) performance in certain (uncertain) environments \cite{Dutta1980,Porter1997}. 
For the hidden action context, it is also shown in Reinwald et al. \cite{Reinwald.2020} that increases in (perceived) environmental turbulence lead to a drop in firm performance. 
This finding is also supported by the contingency management accounting literature, which is, amongst others, concerned with the fit between organizations and their environment (for an overview see, for example, Otley \cite{Otley1980,Otley2016}). If the environment is turbulent and/or the principal is not very well informed about the environment (in terms of the moderating effect of limited intelligence), she cannot design the incentive scheme so that it fits the actual environment \cite{Hoque2005,Chenhall1986,Ezzamel1990,Ghosh2009}. Since the agent responds to the incentives set by the principal, suboptimal incentive parameters directly translate into adverse effects on performance. 

\paragraph{Discussion of results related to the principal's and the agent's utilities}

Above, it was established that the principal's choice of the premium parameter might indicate that she transfers some of the risk to the agent by reducing the premium parameter as environmental turbulence increases. Now, as we take the results related to the agent's utility presented in Sec. \ref{sec:results-utility} into account, this conjecture becomes even more evident. 
One would expect that the agent compensates for the additional risk and the missing risk premium by making less effort, which -- in turn -- eventually increases his utility. However, the results indicate the opposite: The agent's utility decreases with increases in environmental turbulence. If the premium parameter is set by a more intelligent principal, the agent's utility appears to increase. For the principal, the results presented in Sec. \ref{sec:results-utility} indicate that the volatility of his utility is relatively high. However, the principal's utility appears to be robust to limitations in her own as well as in the agent's intelligence. As a consequence -- and this is what can be seen in Fig. \ref{fig:UA_selection} -- one can conclude that the risk-neutral principal withholds a risk-premium from the \textit{risk-averse} agent, which, in turn, assures the robustness of the principal's utility to environmental turbulence.

The results indicate that the agent is {over-dependent} on the principal, which results in a dilemma for the agent: First, if the agent was able to increase his intelligence (in terms of memory), doing so would not allow him to escape the situation, since his own memory has no significant effect on his own utility. Second, the principal appears to set the premium parameter so that the agent is punished for the risky environment. However, this would still be the best possible option for the agent, since otherwise he would have rejected the contract and followed the outside option. Shirking (i.e., putting in less effort) would not be an option either, since it would decrease the agent's utility \cite{Nilakant1994}. Third, even if the agent was more intelligent, he would have no incentive to disclose his private information about the environment. If the principal was informed about the actual exogenous variable (instead of having to estimate it), he could deduce the agent's effort from the outcome \cite{BernardCaillaud.2000}. As soon as the principal realizes that the agent discloses this information, she could simply switch from performance-based pay to effort-based compensation. As a consequence, the principal could further increase her utility at the cost of the agent (see the first best solution in \cite{Holmstrom.1979}). Previous research merely addresses the issue of overly dependent agents and tends to focus on overdependence on the side of the principal: Huang et al. \cite{Huang2016}, for example, state that the principal's power to exert control decreases with the dependence of the agent. Willcocks and Choi \cite{Willcocks1995} also focus on the agent's perspective and put forward the argument that in vendor-client relationships, the client might be overly dependent on the vendor (see also \cite{Hancox2000}). Our results indicate that limited intelligence -- of both the principal and the agent -- appears to empower the principal to (unintentionally) siphon-off utility from the agent by capitalizing on her control over the compensation \cite{David1998}.  

Recall that one fundamental foundation of principal-agent theory is that both the principal and the agent are driven by self-interest \cite{Huang2016}. Now that we know that the principal experiences (almost) the same utility in all cases, she has no incentive to stop transferring risk to the agent or to gather information about the environment on her own (i.e., to increase her intelligence). The principal, thus -- perhaps unintentionally -- behaves in a way that might be interpreted as opportunism, i.e., self-interest-seeking with guile \cite{Williamson1975}, whereby limitations in the principal's intelligence appear to reinforce behavioral patterns that appear as guile.

\subsection{Managerial implications}

From the point of view of managerial implications, it has to be noted that limited intelligence (in terms of limitations in memory) and turbulence might lead to similar effects in the individual perception of the environment. If the principal or the agent are characterized by a minimum level of intelligence -- in our model, this is the case when they take into account the latest estimation/observation only when building their expectation of the environment -- even relatively stable environments might be perceived as being turbulent. Thus, we could conclude that limited intelligence might reinforce the perception of turbulence and weaken the effect of stability in the environment. From the results presented and discussed above, we know that the principal can transfer all negative effects caused by limited intelligence to the agent. She does so by capitalizing on her power over the agent's compensation. Thus, within the context of the model presented in the paper, the principal has no incentive to increase her intelligence, so that her perception of the environment becomes more precise. A constant stream of research in contingency management accounting research asserts that formal controls (i.e., incentive systems such as the one employed in our model) are particularly suited for turbulent conditions \cite{Burns2001,Chenhall2003}. However, there are management controls beyond formal mechanisms, such as interactive- and boundary-based control systems \cite{Simons1995}. It has already been recognized in previous research that less reliance on incentive-based pay, a more subjective evaluation, and a nonaccounting style of performance evaluation might be more efficient in some situations \cite{Moores1998,Bloom1998,Ross1995}. Biased perceptions of the environment -- i.e., the perception of more turbulence than there actually is, which might be moderated by limited intelligence -- might then reinforce the use of rather formal control systems. However, other control mechanisms might be more efficient in such contexts. Consequently, even though management might not have incentives to immediately increase their information processing capabilities (since it appears not to increase their utility), a more precise assessment of the organization's environment might reveal that the entire management control system -- and not only the premium parameter -- needs adaptation to better fit the environment. 

From the perspective of employee satisfaction, transferring risk to the agent might have adverse effects in the long-run. On the one hand, the principal sets the premium parameter so that the agent's utility is higher than what the agent might experience in other employments. In the long-run, however, agents might perceive the organizational culture to be one of exploitation, which, in turn, might lead to a decrease in employee satisfaction and loyalty \cite{Eskildsen2000}. As a result, this might unfold adverse effects on quality and productivity \cite{Juran2003}.\footnote{Regarding the relation between employee satisfaction, loyalty, and performance, the reader is also referred to Silvestro \cite{Silvestro2002}, who provides a more complete review.} In the long-run, agents might even adapt their behavior and shift their priorities so that they might feel inclined to look for another employment. We, therefore, highly recommend to put the results presented in this paper in a long-term context, and not to be seduced by the argument that limited intelligence has no effect on short-term utility, but rather take the long-term effects outlined above into account.

\subsection{Methodological contributions}

From a methodological point of view, we have introduced an agent-based model of the hidden action problem introduced by Holmstr\"om \cite{Holmstrom.1979}. We have employed the agentization approach put forward by Guerrero and Axtell \cite{Guerrero.2011} and Leitner and Behrens \cite{Leitner.2015b}. 

At a conceptual level, our approach is different from the classical principal-agent theory. In particular, agent-based modelling and simulation allow to systematically analyze the robustness of the solutions derived from closed-form models to deviations from the assumptions included in these models \cite{wall2020agent,Guerrero.2011}. The principal-agent theory often includes some idealized assumptions, and some researchers are concerned that (over-)simplified behavioral assumptions might come at the cost of the validity and explanatory power of the findings \cite{Eden1989,Mingers2004,Mingers2011}. Usually it is  assumed that individual behavior is driven by optimization and rationality, the modelled individuals' choices are representative for the entire population, equilibrium solutions can be achieved. Our approach, however, allows for the explicit consideration of emergence, limitations in intelligence, and heterogeneity \cite{Chen2017,Mealy2019,Chang2006}. These features of our approach allow to overcome some of the limitations of the formal approaches in analytical research: While solutions derived from analytical models are related to narrow behavioral assumptions, the model presented here allows to take rich environmental contexts and relaxed -- and perhaps even more realistic -- behavioral assumptions into account \cite{wall2020agent}.  However, the use of agent-based modelling and simulation in microeconomic contexts is rather scarce. Therefore, the approach presented here can be regarded as a step towards a more open approach in microeconomic research that allows for relaxing (some of) the well-established assumptions. 

Our approach also contributes to the stream of behavioral operational research. It is already recognized that developing models that are technically correct and valid might not be sufficient, but it is required to include behavioral insights into the models to inform the efficient design of processes \cite{Franco2020,Royston2013,Hamalainen2013}. It was, amongst others, recognized by White \cite{White2016} that there are a number of scientific fields in which theories of human behavior appear to be becoming more important, such as behavioral decision theory and economics. We merge some developments in research in behavioral control under the umbrella of behavioral operational research. We do so by employing agent-based modelling and simulation at the interface between the fields of economics, management science, operational research, and (cognitive) psychology. We deem our approach particularly relevant in the field of behavioral operational research, since understanding the relationship between knowledge, behavior, and action has at the very core of operational research since the beginning \cite{Ackoff1962}, but it is often asserted that behavioral theory has not yet penetrated the field of operational research \cite{White2016,Bendoly2006}. Thus, from a methodological perspective, we contribute to the further development of behavioral operational research by providing an approach that bridges the involved disciplines' borders.

\section{Conclusions} 
\label{sec:conclusion}

In this paper, we proposed an agent-based model of the hidden action problem. In particular, we transferred the closed-form model introduced in Holmstr\"om \cite{Holmstrom.1979} into an agent-based model by following the agentization approach proposed in Guerrero and Axtell \cite{Guerrero.2011} and Leitner and Behrens \cite{Leitner.2015b}. Doing so allowed us to relax some of the idealized assumptions in Holmstr\"om's model that are related to the principal's and the agent's respective intelligence. We put a particular focus on the memory related to information about the environment. Our results indicate that the principal's intelligence is the key to good performance, whereas the agent's intelligence appears not to affect performance significantly. Surprisingly, the principal appears to behave in a very selfish way by siphoning off utility from the agent to maintain near-optimal personal utility by exerting her control over the agent's compensation. Our research contributes to an important stream of behavioral operational research. We model (more realistic) human behavior by employing the methods of operational research and by considering (and bringing together9 findings from the disciplines of cognitive psychology, economics, management, and operational research. 

Of course, our research is not without limitations. As already discussed in Sec. \ref{sec:discussion}, there is a multiplicity of further incentive mechanisms -- also nonformal ones -- that might be employed in the modelled situation. Granting the principal degrees of freedom in her choices related to the control mechanism might be a fruitful avenue for future research. We limit the principal's and the agent's respective intelligence with respect to memory only. Further research might focus on extending the limitations in intelligence by also taking into account, for example, calculation errors, limitations in other types of information, and biases in information processing.

%\section*{Conflict of interest}
%\noindent The authors declare that they have no conflict of interest.

%\section*{Author contributions}
%\noindent \textbf{Patrick Reinwald:} Conceptualization, Methodology, Software, Validation, Formal analysis, Data curation, Writing, Visualization.
%
%\textbf{Stephan Leitner:} Conceptualization, Methodology, Software, Validation, Formal analysis, Data curation, Writing, Visualization, Supervision, Project administration, Funding acquisition.
%
%\textbf{Friederike Wall:} Conceptualization, Methodology, Formal analysis, Writing, Visualization, Supervision, Funding acquisition.

%\section*{Role of the funding source}
%\noindent The funding source had \textit{no involvement} in study design, the analysis and interpretation of data, in the writing of the manuscript, and in the decision to submit the article for publication.

\clearpage
\setcounter{table}{0}
\setcounter{figure}{0}

\appendix

\counterwithin{figure}{section}

\section{Results of the simulation study}
\label{Appendix_A}

This appendix includes the full set of results of the simulation study. The considered simulation parameters are included in the following table:
\begin{table}[ht]
\small
	\label{tab:key_para-appendix}   
	\begin{tabular}{lll}
		Parameter & Notation & Values \\
		\noalign{\smallskip}\hline\noalign{\smallskip}
		\textbf{Subject to variation}\\
		Principal's memory & $ m_P $ & $\{1,3,5,\infty\}$\\
		Agent's memory & $ m_A $  & $\{1,3,5,\infty\}$\\
		Exogenous factor: standard deviation & $ \sigma $ & $\{0.05x^{\ast},0.25x^{\ast}, 0.45x^{\ast}\}$\\
		\textbf{Constants}\\
	    Exogenous factor: mean & $ \mu $ & $ 0$\\
		Agent's Arrow-Pratt measure & $ \eta $ & $0.5$\\
		Observation period  &   $T$ &   $20$ \\
		Simulation rounds  &   $R$ &   $700$ \\
		\noalign{\smallskip}\hline
	\end{tabular}
		\caption{Simulation parameters}
\end{table}

\begin{figure}[ht]
	\centering
	\includegraphics[width=\textwidth]{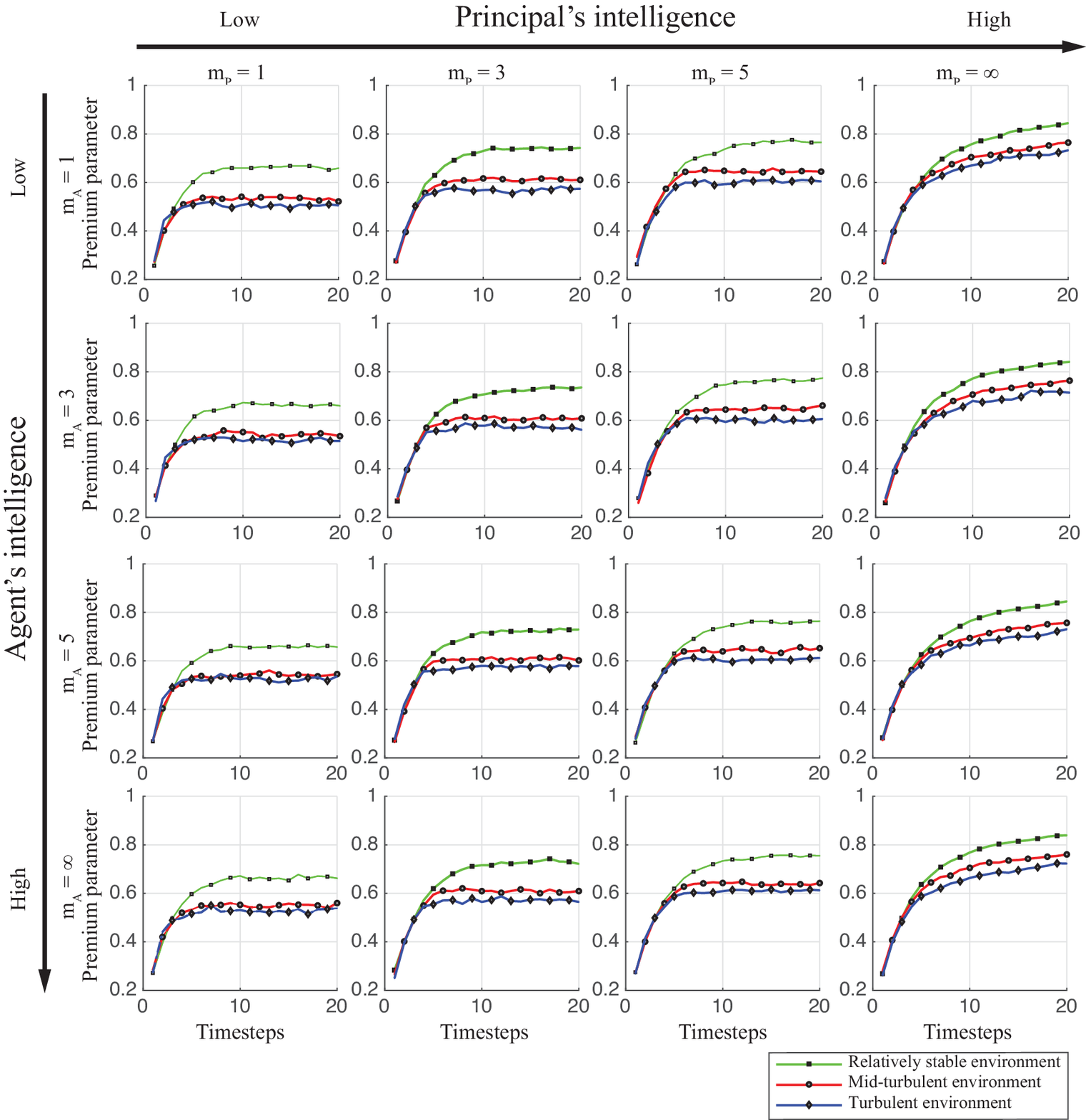}
	\caption{Normalized average premium parameter for all scenarios. Shaded areas represent confidence intervals at the 99\%-level}
	\label{fig:premiump_all}
\end{figure}

\begin{figure}[ht]
	\centering
	\includegraphics[width=\textwidth]{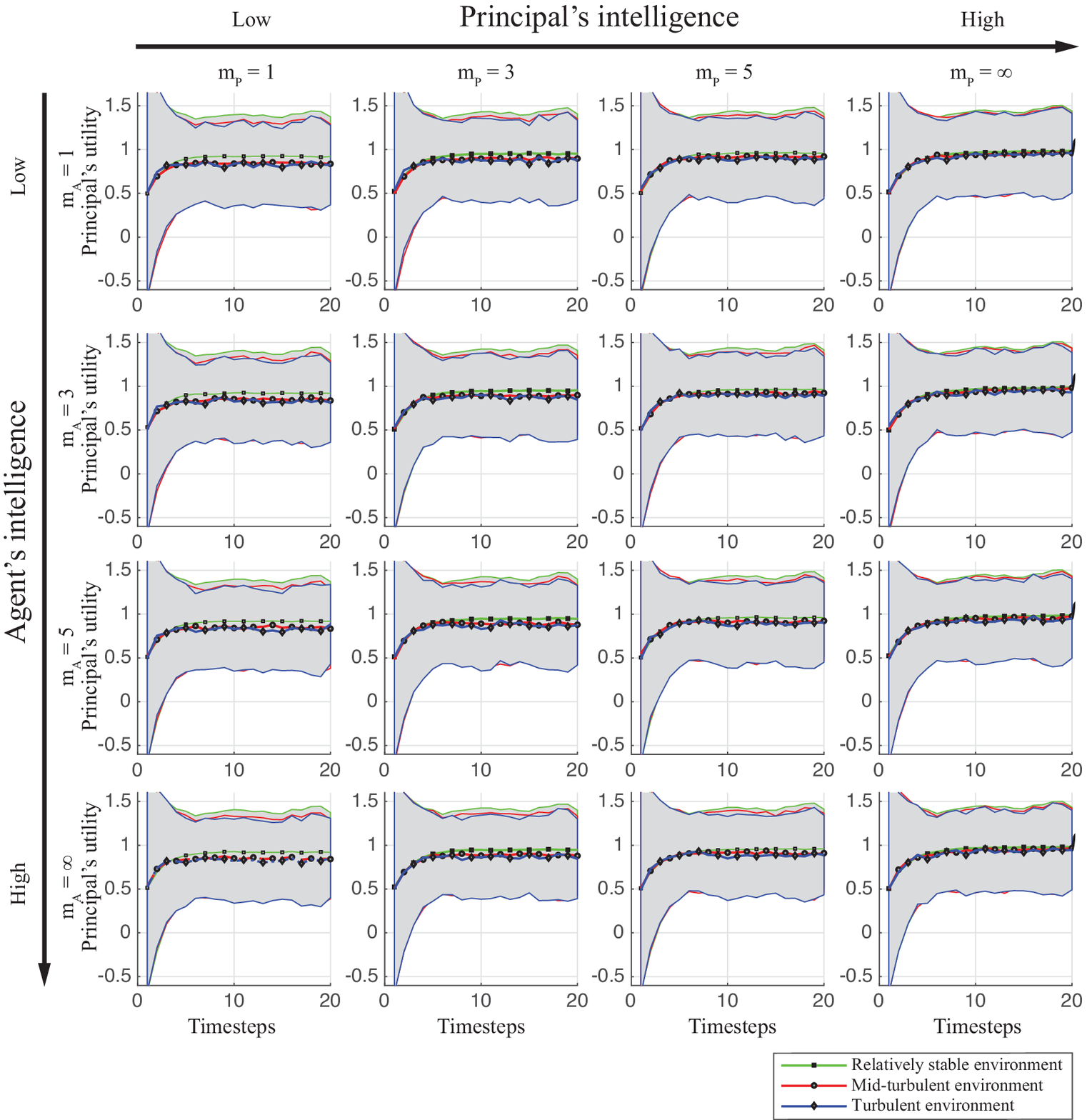}
	\caption{Principal's normalized average utility for all scenarios. Shaded areas represent confidence intervals at the 99\%-level}
	\label{fig:UP_all}
\end{figure}

\begin{figure}[ht]
	\centering
	\includegraphics[width=\textwidth]{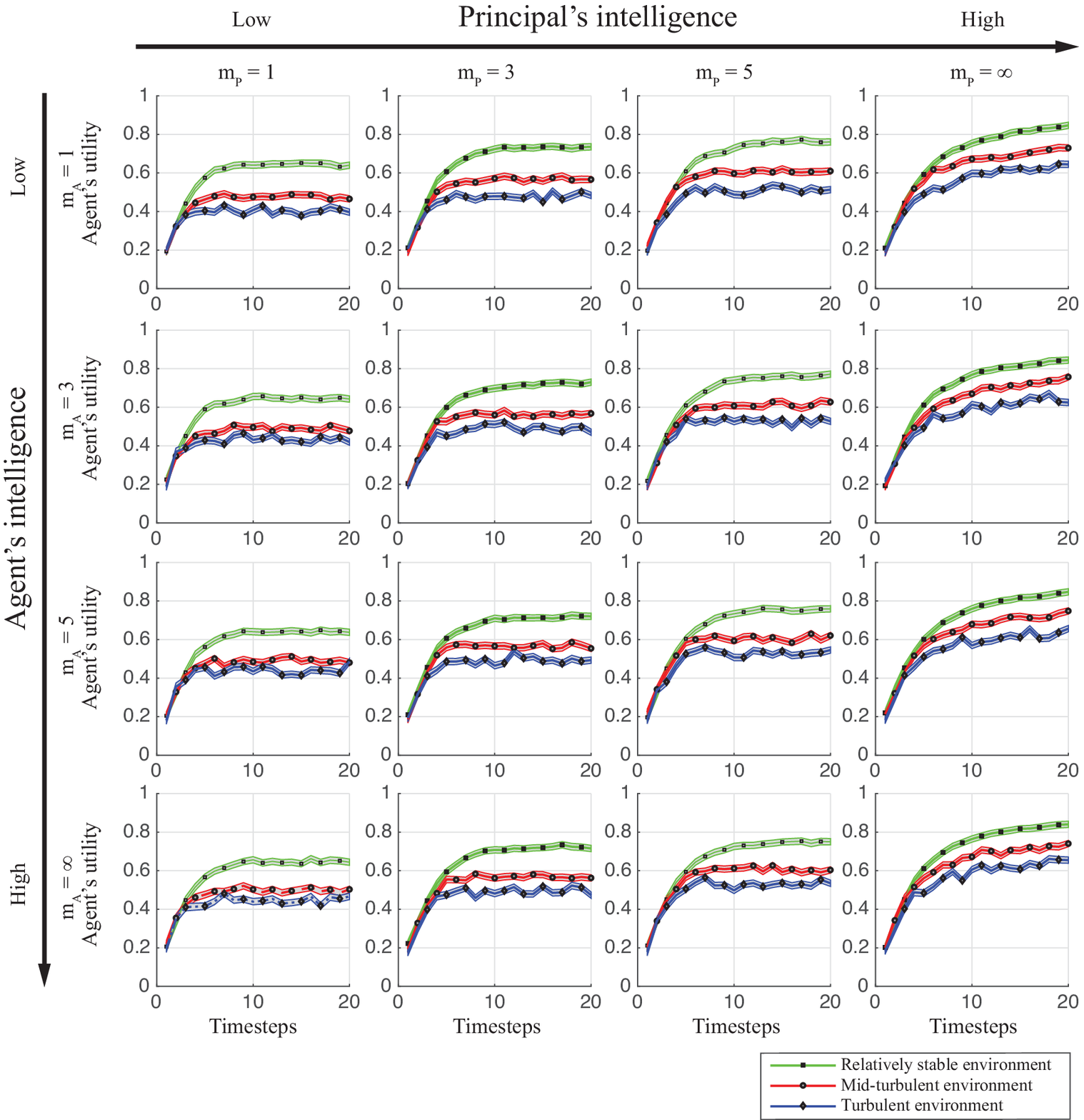}
	\caption{Agent's normalized average utility for all scenarios. Shaded areas represent confidence intervals at the 99\%-level}
	\label{fig:UA_all}
\end{figure}

\begin{figure}[ht]
	\centering
	\includegraphics[width=\textwidth]{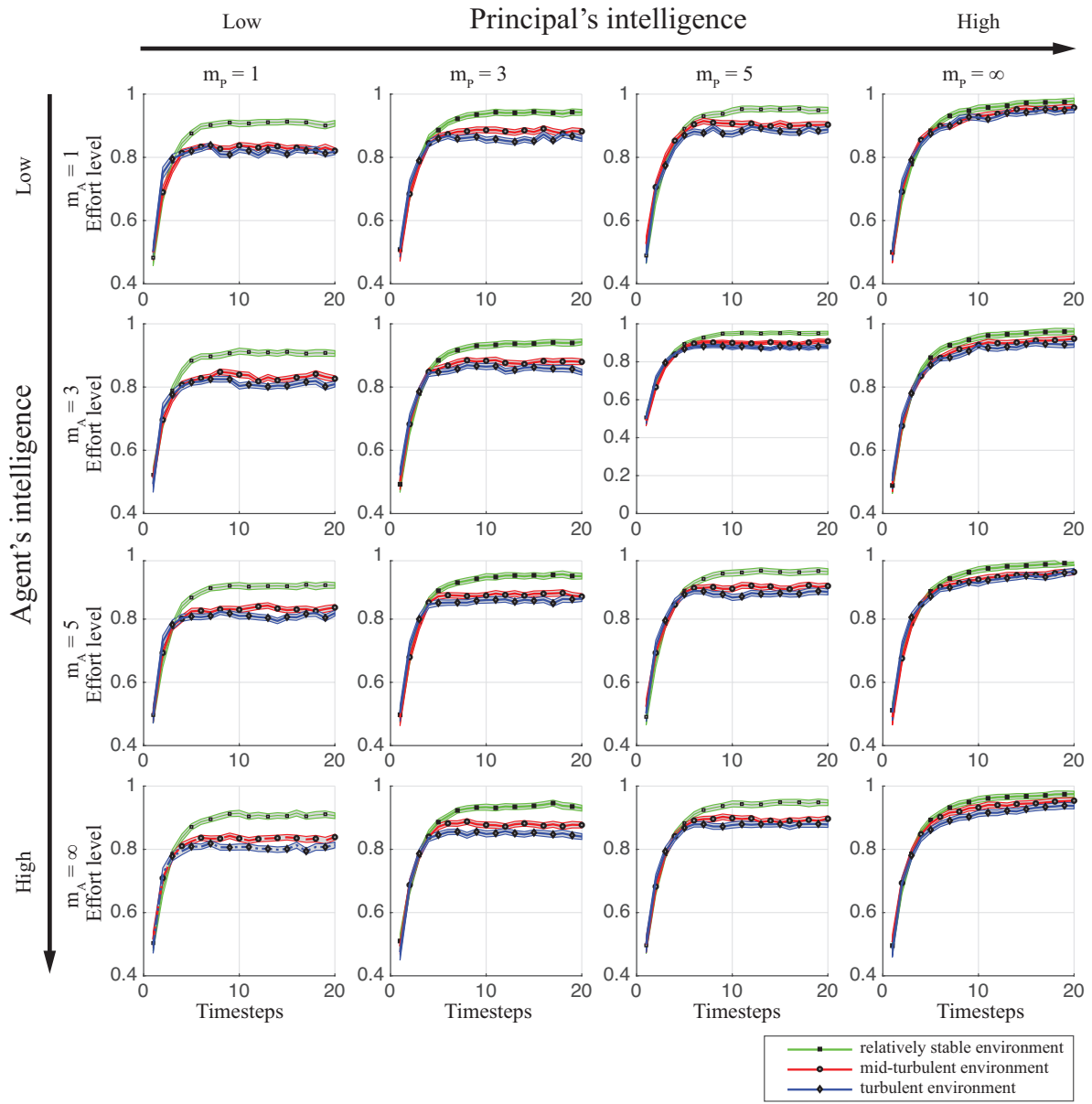}
	\caption{Normalized average effort for all scenarios. Shaded areas represent confidence intervals at the 99\%-level}
	\label{fig:a_A_all}
\end{figure}

\clearpage

\section{Solution to Holmstr\"om's hidden action problem}
\label{Appendix_B}

There are two different approaches that can be used to solve the program formalized in Eqs.  \ref{eq:begin}-\ref{eq:incenti}. Important for us is the approach of Mirrlees (1974), who suppresses $ \theta $ and views $ x $ as a random variable with distribution $ F(x,a) $. In this approach, it is assumed that $ \forall \: a \in A \: \exists \: x \in {\rm I\!R} : F_a(x,a) < 0 $ so that a change in $ a $ has nontrivial effect on the distribution of $ x $. For a given distribution of $ \theta $, $ F(x,a) $ is the distribution induced on $ x=x(a,\theta) $ \cite{Holmstrom.1979}. 

In the following program, $ f(x,a) $ is the density function of $ F $ with $ f_a $ and $ f_{aa} $ well defined for all $ (x,a) $ and Eq. (\ref{eq:incenti}) is replaced with a first-order constraint. Furthermore, $ s(x) $ is restricted to lie in the interval $ [c,d+x] $ to guarantee an existing solution to Eqs. \ref{eq:begin}-\ref{eq:incenti} for the class of functions in Eq. \ref{eq:classFun}, where $ V^{b'}_{b} $ is the total variation of $ s $ in the interval $ [b,b'] $ \cite{Kolmogorov.1970,Holmstrom.1979}. 

\begin{equation}
\label{eq:classFun}
S_K = \{s(x)\in[c,d+x]|V^{b'}_{b}(s)\leq K\cdot(b'-b)\},
\end{equation}

\begin{subequations}
	\begin{eqnarray}
	\label{eq:soleins}
	\underset{s(x) \in[c,d+x],a}{max}\int G(x-s(x))f(x,a)dx 
	\end{eqnarray}
	\begin{eqnarray}
	\label{eq:solzwei}
	subject \:to \int [U(s(x)-V(a)]f(x,a)dx \geq \bar{H},
	\end{eqnarray}
	\begin{eqnarray}
	\label{eq:soldrei}
	\int U(s(x))f_a(x,a)dx = V'(a).
	\end{eqnarray}
We denote the multipliers for Eqs. \ref{eq:solzwei} and \ref{eq:soldrei} by $\lambda$ and $\mu $, respectively. 
After a pointwise Lagrangian optimisation, the optimal sharing rule yields the following characterization:
\end{subequations}
\begin{equation}
\label{eq:solvier}
\dfrac{G'(x-s(x))}{U'(s(x))} = \lambda + \mu \cdot \dfrac{f_a(x,a)}{f(x,a)},
\end{equation}
for almost every x for which Eq. \ref{eq:solvier} has a solution $ s(x) \in [c,d+x] $.
Also, $\mu$ is given as solution to the adjoint equation and is determined by Eq. \ref{eq:soldrei} \cite{Holmstrom.1979}.
\begin{equation}
\label{eq:fun}
\int G(x-s(x))f_a(x,a)dx+\mu\{\int U(s(x))f_{aa}(x,a)dx-V''(a)\}=0
\end{equation}

\clearpage

\bibliography{bibio.bib}

\end{document}